\documentclass[12pt,a4wide]{article} % use larger type; default would be 10pt

\usepackage[utf8]{inputenc} % set input encoding (not needed with XeLaTeX)

\usepackage{amsmath}
\usepackage{amsfonts}
\usepackage{amssymb}
\usepackage{graphicx}%
\numberwithin{equation}{section}

\usepackage{geometry} % to change the page dimensions
\usepackage{graphicx} % support the \includegraphics command and options

\usepackage{booktabs} % for much better looking tables
\usepackage{array} % for better arrays (eg matrices) in maths
\usepackage{paralist} % very flexible & customisable lists (eg. enumerate/itemize, etc.)
\usepackage{verbatim} % adds environment for commenting out blocks of text & for better verbatim
\usepackage{subfig} % make it possible to include more than one captioned figure/table in a single float
\usepackage{times}

\usepackage[usenames,dvipsnames]{color}

\usepackage{hyperref}
\hypersetup{
colorlinks=true,
citecolor=Brown,
linkcolor=blue
}

\usepackage{fancyhdr} % This should be set AFTER setting up the page geometry
\usepackage{sectsty}
\allsectionsfont{\sffamily\mdseries\upshape} % (See the fntguide.pdf for font help)
\usepackage[nottoc,notlof,notlot]{tocbibind} % Put the bibliography in the ToC
\usepackage[titles,subfigure]{tocloft} % Alter the style of the Table of Contents

\newcommand{\ignore}[1]{}

\newcommand{\etal}{\textit{et al}.}
\DeclareRobustCommand{\mm}[1]{}
\renewcommand{\texttt}{\mm}

\newcommand{\hatn}{\hat{n}}

\title{\textbf{Gauge gravity vacuum in constraintless Clairaut-type formalism}}
\author{\textit{\bfseries Michael L. Walker}\\
University of New South Wales, Kensington, NSW 2033, Australia\\
 and\\ \textit{\bfseries Steven Duplij}\\
 Center for Information Technology (WWU IT),
University of M\"unster,\\
R\"ontgenstrasse 7-13,
D-48149 M\"unster,
Germany}
\date{\small v. 6: January 11, 2022}

\begin{document}

\maketitle
\begin{abstract}
%\noindent
The gauged Lorentz theory with torsion has been argued to have an effective theory whose non-trivial background is responsible for background gravitational curvature if torsion is treated as 
a quantum-mechanical variable against a background of constant curvature. We use the CDG decomposition to argue that such a background can be found without including torsion. 
Adapting our previously published Clairaut-based treatment of QCD, we go on to study the implications for second quantisation.
\end{abstract}
\thispagestyle{empty}
\newpage
\mbox{}

\bigskip

\begin{small}
\tableofcontents
\end{small}
%\newpage

\bigskip

\section{Introduction}
The dual superconductor model of QCD confinement requires the vacuum to contain a condensate of (chromo)
magnetic monopoles. This led several authors to consider embedded, usually Abelian, subgroups within gauge groups.
The early focus was on the $U(1)$ subgroup of $SU(2)$, with analyses by Savvidy \cite{S77}, Nielsen and Olesen \cite{NO78} and t'Hooft \cite{tH81} 
considering the maximal Abelian gauge in which the Abelian subgroup is assumed to lie along the internal $e_3$-axis. While they did find a 
magnetic condensate to be a lower energy state than the perturbative vacuum, their analyses blatantly violated gauge covariance and offered no evidence that the chromomagnetic background
was due to monopoles. There was also considerable controversy regarding the stability of such a vacuum.
These issues were resolved by the Cho-Duan-Ge (CDG) decomposition \cite{Cho80a, DG79} 
which introduces an internal vector to covariantly allow a subgroup embedding within a theory's gauge group to vary throughout spacetime.
Analyses based on this approach confirmed this magnetic background \cite{S77, tH81} and careful consideration of renormalisation and causality \cite{CP02,Cme04,CmeP04,KKP05}
finally resolved such a condensate to be stable through several independent arguments.

It is common for analyses of QCD based on the CDG decomposition to assume 
the monopole condensate comprising the vacuum to provide a slow-moving vacuum
background to the quantum degrees of freedom (DOFs) \cite{CP02}. This was the basis of a novel approach to 
Einstein-Cartan gravity, in which contorsion (or torsion) is the quantised dynamic degree of freedom confined by a slow-moving classical background gravitational 
curvature \etal~\cite{KP08,P10,CPP10}.
Their work was based on the Lorentz gauge field theory initially put forward by Utiyama-Kibble-Sciama \cite{U56,K61,S64} for which it has long been known that the non-compact 
nature of the Lorentz group led to the theory not being postive semi-definite. They dealt with this by performing their initial analyses in Euclidean space, transforming 
the Lorentz gauge group to $SO(4) \simeq SU(2) \times SU(2)$, until later work found the theory to be well-defined with propagators for its canonical DOFs \cite{PKT12}
%Indeed, several arguments have been given against treating the metric as the fundamental field, such as the necessity of interacting with fermion fields.
%Their work includes an analysis of the contorsion DOFs which finds the general form of the Lagrangian for which the contorsion has a well-defined propagator.
Instead of including contorsion, we consider the Abelian decomposition of the Lorentz gauge field strength. Drawing on 
a considerable body of literature concerning the Abelian decomposition of $SU(2)$ Yang-Mills, 
we find an interesting structure without the introduction of contorsion. 

To avoid third order derivatives from entering the equations of motion (EOMs), our theory does not include localised translation symmetry, 
although it is accepted that spacetime respects the full Poincar\'{e} symmetry group. 
We restrict ourselves to the subgroup in this work to avoid complications and so that we can 
find conventional propagators for the gauge bosons with a Lagrangian quadratic in gravitational curvature. 
We remain mindful, however, that this is a reduced symmetry group of gravitational dynamics rendering our model to be either low-energy
effective or perhaps even just a toy.
%which illustrates the richness of even the simplest quantised approaches to gravity. 
%Indeed, we find strong parallels between our consideration of gauged Lorentz symmetry 
%and the role played by $SU(2)$ allows the techniques previously applied to %earlier analyses of the vacuum of
%two-colour QCD, and some of their results, to be applicable here also. 
%In particular, which allow us to adapt results concerning the stability of that vacuum to this theory. 

%We do not, therefore, consider it a full theory of gravity, but rather a toy model for providing insight.

One of the more confusing mathematical subtleties of the CDG decomposition was the number of canonical degrees of freedom. 
Shabanov argued that an additional gauge-fixing condition is needed to remove a supposed \textquotedblleft two extra degrees\textquotedblright \cite{S99b}
introduced by the internal unit vector field to covariantly describe the embedded subgroup(s). Bae, Cho and Kimm later clarified that 
this internal vector did not introduce two degrees of freedom requiring to be fixed but non-canonical DOFs without EOMs \cite{BCK02},
while the proposed constraint was merely a consistency condition.
The interested reader is referred to
\cite{S99,CP02,KMS05,K06} for further details (see, also \cite{lav/mer2016,ren/wan/qu}).
Cho \etal{} \cite{CHKP07} approached the issue with Dirac quantisation using second-order restraints.
In an earlier paper \cite{meD15} the authors however a new approach to
rigorously elucidate the dynamic DOFs from the topological. It is based on the Clairaut-type formulation, proposed by one of the authors (SD)
\cite{D10,D14}, in a constraintless generalization of the standard Hamiltonian formalism to include Hessians with zero determinant. It provides
a rigorous treatment of the non-physical DOFs in the derivation of EOMs and the quantum commutation relations.
In this paper we apply our Clairaut approach to the gauged Lorentz group  \cite{COK12,COP15} theory with a Lagrangian quadratic in curvature. 

A review of the CDG decomposition is given in Section \ref{sec:A2}, beginning with an introduction in the context of QCD before illustrating its application to 
$SU(2)\times SU(2)$. In Section \ref{sec:parallel} we illustrate the reduction of our theory to two copies of two-colour QCD
and use one-loop results from the latter to inform us about the former. Section \ref{sec:Clairaut} gives 
a brief overview of the Clairaut-Hamiltonian formalism and uses it to study the quantisation of this theory, 
sorting canonical dynamic DOFs from DOFs describing the embedding of important subgroups and finding deviations from canonical second quantisation even for dynamic fields. 
We consider the one-loop effective dynamics in Section \ref{sec:effective}, discussing the effective
particle spectrum in Subsection \ref{subsec:operator} and the possible emergence of the Einstein-Hilbert (EH) term in Subsection \ref{subsec:EH}.
Our final discussion is in Section \ref{sec:discussion}.

\section{A review of the covariant Abelian decomposition of gravity} \label{sec:A2} %The covariant specification of the Abelian dynamics of gravity}
\subsection{\label{subsec:CDG}The CDG decomposition in $SU(2)$ QCD}
\subsubsection{Formalism}
Abelian dominance has played a major role in our understanding of the QCD vacuum, facilitating the demonstration of a monopole condensate.
That a magnetic condensate suitable for colour confinement can have lower energy than the perturbative vacuum has been known since the 1970s \cite{S77,NO78,tH81}, but in early work the internal direction 
supporting the magnetic background could not be specified in a covariant manner and nor was there support for the magnetic condensate 
being due to monopoles. The apparent existence of destabilising tachyon modes was also an issue for some time \cite{NO78,S82,CmeP04}. 
These issues were rectified by the introduction of the CDG decomposition, which 
specifies the internal direction of the Abelian subgroup in a gauge covariant manner,
allowing the internal direction to vary arbitrarily throughout spacetime.

The application of the CDG decomposition in $N$-colour ($SU(N)$) QCD is as follows:\newline
%the CDG decomposition, as per the convention of Cho \emph{et al.} \cite{CGZ12}.
The Lie group $SU(N)$ has $N^{2}-1$ generators $\lambda^{(a)}$ ($a=1,\ldots
N^{2}-1$), of which $N-1$ are Abelian generators $\Lambda^{(i)}$ ($i=1,\ldots
N-1$).

The gauge transformed Abelian directions (Cartan generators) are denoted as%
\begin{equation}
\hat{n}_{i}(x)=U(x)^{\dagger}\Lambda^{(i)}U(x).
\end{equation}

Gluon fluctuations in the $\hat{n}_{i}$ directions are described by
$c_{\mu}^{(i)}$, where $\mu$ is the Minkowski index. There is a covariant
derivative which leaves the $\hat{n}_{i}$ invariant,
\begin{equation}
\label{eq:Dhat}\hat{D}_{\mu}\hat{n}_{i}(x)\equiv(\partial_{\mu}+g\vec{V}_{\mu
}(x)\times)\hat{n}_{i}(x)=0,
\end{equation}
where $\vec{V}_{\mu}(x)$ is of the form
\begin{equation}
\label{eq:vecV}\vec{V}_{\mu}(x)=c_{\mu}^{(i)}(x)\hat{n}_{i}(x)+\vec{C}_{\mu
}(x), \quad\vec{C}_{\mu}(x)=g^{-1}\partial_{\mu}\hat{n}_{i}(x)\times\hat
{n}_{i}(x).
\end{equation}
The vector notation refers to the internal space, and summation is implied
over $i=1,\ldots N-1$. For later convenience we define
\begin{align}
F^{(i)}_{\mu\nu}(x) = \partial_{\mu}c^{(i)}_{\nu}(x)- \partial_{\nu}%
c^{(i)}_{\mu}(x),& \\
\vec{H}_{\mu\nu}(x) = \partial_{\mu}\vec{C}_{\nu}(x)- \partial_{\nu}\vec
{C}_{\mu}(x) +g\vec{C}_{\mu}(x)\times\vec{C}_{\nu}(x)= &\partial_\mu \hatn_i(x) \times \partial_\nu \hatn_i(x), \label{eq:H}  \\	
% = H^{(i)}_{\mu\nu}(x) \hat{n}_{i}(x), \nonumber \\ 
H^{(i)}_{\mu\nu}(x) = \vec{H}_{\mu\nu}(x) \cdot\hat{n}_{i}(x),& \\
\vec{F}^{(i)}_{\mu\nu}(x) = F^{(i)}_{\mu\nu}(x) \hatn_i(x) + \vec{H}_{\mu\nu}(x)&.
\end{align}
The second last term in eqn~(\ref{eq:H}) follows from the definition in eqn~(\ref{eq:vecV}). Its being a cross-product is significant as it prevents $\mu,\nu$ from having the same value.
The Lagrangian contains the square of this value, namely
\begin{equation}
H^{(i)}_{\mu\nu}(x) H_{(i)}^{\mu\nu}(x) = \left(\partial_\mu \hatn_i(x) \times \partial_\nu \hatn_i(x)\right) \cdot \left(\partial^\mu \hatn_i(x) \times \partial^\nu \hatn_i(x)\right), 
\end{equation}
The form of eqn~(\ref{eq:vecV}) might suggest the possibility of third or higher time derivatives in a quadratic Lagrangian, but we have now seen that the specific form of the Cho connection
does not allow this.

The dynamical components of the gluon
field in the off-diagonal directions of the internal space vectors are denoted by $\vec{X}_{\mu}(x)$, so if $\vec{A}_{\mu}(x)$ is the gluon field then%
\begin{equation}
\vec{A}_{\mu}(x)=\vec{V}_{\mu}(x)+\vec{X}_{\mu}(x)=c_{\mu}^{(i)}(x)\hat{n}%
_{i}(x) +\vec{C}_{\mu}(x)+\vec{X}_{\mu}(x),
\end{equation}
where
\begin{equation}
\vec{X}_{\mu}(x) \bot\hat{n}_{i}(x),\; \forall\, 1\le i<N\,,\quad\vec{D}_{\mu}
=\partial_{\mu}+g\vec{A}_{\mu}(x).
\end{equation}

The Lagrangian density is still
\begin{equation} \label{eq:QCD}
\mathcal{L}_{gauge}(x) = -\frac{1}{4} \vec{R}_{\mu\nu}(x) \cdot\vec{R}^{\mu\nu}(x),
\end{equation}
where the field strength tensor of QCD expressed in terms of the CDG
decomposition is
\begin{align} \label{eq:fieldstrength}
\vec{R}_{\mu\nu}(x)  &  =\vec{F}_{\mu\nu}(x) +(\hat{D}_{\mu}\vec{X}_{\nu}(x)-\hat{D}_{\nu}\vec{X}_{\mu}(x)) +g\vec{X}_{\mu}(x)\times\vec{X}_{\nu}(x).
\end{align}

Gauge transformations are effected with a gauge parameter $\vec{\alpha}(x)$.
Under a gauge transformation $\delta$ with $SU(2)$ parameter $\vec{\alpha}(x)$
\begin{align} \label{eq:transform}
&\delta \hat{V}(x) =  \, \hat{D}_\mu \vec{\alpha}(x) \nonumber \\
&\delta c_\mu(x) =  \,(\partial_\mu \vec{\alpha}(x) \cdot \hatn(x)), \nonumber \\
&\delta \hatn(x) = \,\hatn(x) \times \vec{\alpha}(x), \nonumber \\
&\delta \vec{C}_\mu(x)=  \,(\partial_\mu \vec{\alpha}(x))_{\perp\hatn} + g\vec{C}_\mu(x) \times \vec{\alpha}(x), \nonumber \\
&\delta \vec{X}_\mu(x) =  \,g \,\vec{X}_\mu(x) \times \vec{\alpha}(x).
\end{align}
The form of the transform for $\vec{X}_\mu$ is the same as that for a coloured source, so that these components are sometimes described as \textquotedblleft valence\textquotedblright. This gauge transformation tell us two interesting things. The first is that the Abelian component $c_\mu$ combined with the Cho connection
$\vec{C}_\mu$ are enough to represent the full Lorentz symmetry even without the valence components $\vec{X}_\mu$, Cho \etal~\cite{COK12,COP15} described as the "restricted"  theory. 
The second is that the valence components transform like 
a source transforms. There is a corresponding situation in $N=2$ Yang-Mills where the valence gluons are interpreted as colour sources. The importance of this observation 
is that we shall later discuss the possibility of mass generation for the valence gluons and this form for the gauge transformation leaves such mass terms covariant. We note however that 
a bare mass for $\vec{X}_\mu$ cannot be inserted artificially without spoiling renormalisability.

\subsubsection{The degrees of freedom in the CDG decomposition} \label{subsubsec:DOF}
Henceforth we restrict ourselves to the $SU(2)$ theory, for which there is
only one $\hat{n}$ lying in a three dimensional internal space, and neglect
the $(i)$ indices. 

The unit vector $\hatn$ posseses two DOFs and so its inclusion in the gluon field together with the Abelian component $c_\mu$ and 
the valence gluons $\vec{X}_\mu$ raises questions about the DOF of the decomposed gluon, with one paper \cite{S99b} advocating the gauge condition
\begin{equation} \label{eq:XfixQCD}
\hat{D}_\mu \vec{X}_\mu(x) = 0,
\end{equation}
to remove two apparent 
extra degrees of freedom. The matter was sorted by Bae \etal~\cite{BCK02} who demonstrated that the DOFs of $\hatn$ were not canonical but topological,
indicating the embedding of the Abelian subgroup in the gauge group. The canonical DOFs are carried by the components $c_\mu, \vec{X}_\mu$ and eqn~(\ref{eq:XfixQCD}) is 
a consistency condition expected of valence gluons. Kondo \etal{}~\cite{KMS06} considered a stronger condition guaranteed not to be unaffected by Gribov copys.

The topological nature of $\hatn$ has significance beyond making the canonical DOFs add up correctly. As is well known, monopole configurations in gauge theories are topological 
configurations corresponding to the embedding of an Abelian subgroup. The other important consequence is that $\hatn$ does not have 
a canonical EOM from the Euler-Lagrange equation. 
%Attempts have been made to derive an EOM for the monopole field using Dirac quantisation and second class constraints \cite{CHKP07}.

We took an alternative approach to this issue by applying a new method for finding the effects of degenerate variables called the Clairaut formalism.
We further assumed that, as a unit vector, its dynamics were best described by angular variables.

\subsection{CDG decomposition of $SU(2) \times SU(2)$ in Euclidean space} \label{subsec:labelEuclid}
As is well known \cite{KP08,PKT12,U56,K61,S64}, the non-compact nature of the Lorentz group causes Lorentz gauge theories to be non-positive semi-definite. In fact, our attempts to 
apply the CDG decomposition to the Lorentz gauge field strength tensor in Minkowski space led to negative kinetic energy terms for some of the gauge fields 
(not shown). As demonstrated by Pak \etal{}~\cite{KP08,PKT12}, this can be avoided by Wick rotating the theory to Euclidean space and then either 
considering effective theories or finding a way to rotate back later without spoiling the quantum theory.

This procedure also rotates the internal Lorentz group to $SO(4)$ which is locally isomorphic to 
$SU(2)_R \times SU(2)_L$, corresponding to the right- and left- handed groups generated by 
\begin{equation} \label{eq:generators}
\frac{1}{\sqrt{2}}\left(J_l \pm i K_l\right),
\end{equation}
where $J_l, K_l$ are the 
rotation and boost operators, respectively.
The two $SU(2)$ subgroups in our gauge theory, though separate, are not independent but are built from the same rotation and boost operators, albeit in
combinations of opposite chirality. 
It follows that their respective Abelian directions must correspond, but represent operators of different chirality. 
%requiring the 
%introduction of only one $\hatn$ field which must apply to both $SU(2)$ subgroups. The CDG decomposition now parameterises the symmetry reduction 
%of the gauge group to $U(1) \times U(1)$ where both $U(1)$ are indicated by the same unit vector $\hat{n}$.
We denote them $\hatn_R, \hatn_L$ respectively, using these suffices for other field objects also when appropriate, and apply previously published analyses \cite{S77,tH81,CP02,Cme04,CmeP04} 
to each symmetry group.

We apply the CDG decomposition to $SU(2)_R \times SU(2)_L$ gauge group. 
Their Abelian components we denote $_Rc_\mu$ and $_L{c}_\mu$ respectively and the valence components we denote as
$_R\vec{X}_\mu$ and $_L\vec{X}_\mu$ respectively. For each chirality $\chi \in \{R,L\}$ we have the Cho connection
\begin{equation} \label{eq:connection}
_\chi C_\mu(x) = g^{-1}\partial_{\mu} \,_\chi\hatn(x) \times {}_\chi\hatn(x),
\end{equation}
and monopole field strength
\begin{align} %\label{eq:H}
_\chi \vec{H}_{\mu\nu}  \equiv \partial_\mu {}_\chi\vec{C}_\nu(x) - \partial_\nu {}_\chi\vec{C}_\mu(x) + g \, {}_\chi\vec{C}_\mu(x) \times {}_\chi\vec{C}_\nu(x) 
&= \partial_\mu \, \hatn_\chi(x) \times \partial_\nu \,\hatn_\chi(x) \nonumber \\
&\equiv {}_\chi H_{\mu\nu}(x) \,\hatn_\chi(x). 
\end{align}

\section{\label{sec:parallel}The vacuum of  $SU(2)_R \times SU(2)_L$}
Since the component $SU(2)$ symmetry groups have generators mutually orthogonal in the internal space their contributions to the ground state may be calculated independently 
and summed. Furthermore, their identical fundamental dynamics imply that $_\chi H_{\mu\nu}$ is independent of $\chi$ when we are not considering an internal vector 
and may be replaced with $H_{\mu\nu}$, which we do henceforth.

It is sufficient to calculate to one loop to find a non-zero monopole condensate in the effective action of $SU(2)$ Yang-Mills theory. References \cite{CP02,Cme04,CmeP04} have shown this by 
a variety of methods. Useful material on this theory at one-loop order can also be found in references  \cite{IZbook12,PSbook95,Wbook96}.

Calculating the relevant one-loop Feynman diagrams iin Feynman gauge with dimensional regularisation \cite{Cme04,CmeP04} we have
\begin{equation}
\Delta S_{eff} = -\frac{11g^2}{96} \sum_{\chi=R,L} \int d^4p\,_\chi\vec{F}_{\mu\nu}(p)\,_\chi\vec{F}_{\mu\nu}(-p) \left( \frac{2}{\epsilon} - \gamma - \ln\Big(\frac{p^2}{\mu^2}\Big)\right).
\end{equation}
An imaginary part is generated by the $\ln\frac{p^2}{\mu^2}$ term only when the momentum $p$ is timelike, leading to the well-known result \cite{S51,Cme04,CmeP04} 
that it is the electric backgrounds are unstable 
but magnetic ones are not. Using this information we then have the effective potential 
\begin{equation}
V = \frac{H^2}{g^2}
\left[1 + \frac{11g^2}{24} \Big(\ln\frac{\sqrt{H^2}}{\mu^2} - c\Big)\right]
\end{equation}
It should be remembered that this close parallel with the corresponding $N=2$ calculation does not hold beyond one loop because then there are diagrams including fields from both $SU(2)$ subgroups.

Defining the running coupling $\bar{g}$ by \cite{Cme04,CmeP04}
\begin{equation}
\frac{\partial^2V}{\partial H^2} \Big|_{\sqrt{H^2}=\bar{\mu}^2}
=\frac{1}{\bar{g}^2} .
\end{equation}
leads to a non-trivial local minimum at
\begin{equation}
\langle H \rangle = \bar{\mu}^2 \exp\Big(-\frac{24\pi^2}{11\bar{g}^2} + 1\Big).
\end{equation}

The specific value of $H^2$ is less important than knowing it has a strictly positive value lying in two orthogonal directions in the $SU(2)_R\times SU(2)_L$ internal space. 
\ignore{
Remembering the form of the cross product in eqn (\ref{eq:cross}), the three point interactions involving the Lorentz-monopole component are limited to the sets $\{C_\mu,c_\nu,\vec{X}_\lambda\}$
and $\{C_\mu(x),\tilde{c}_\nu(x),\vec{X}_\lambda(x)\}$ plus those involving ghost terms, which is clearly just a duplicate of the three-point vertices in the CDG decomposition of $N=2$ Yang-Mills 
theory (see Section \ref{subsec:CDG} or references \cite{Cme04,CmeP04}). This is almost the case also for four-point interactions involving the monopole component, except for an additional vertex for
the set $\{C_\mu(x),\vec{X}_\nu(x),\vec{X}_\lambda(x)\}$. This clearly indicates that differences between the two theories will emerge, but not at
one-loop level. We may, therefore, consider effects from this theory at one-loop to be double those from Yang-Mills to this theory up to one-loop level, but must be careful beyond that.

Working through the same one-loop calculations as in the analysis of $SU(2)$ Yang-Mills theory \cite{Cme04,CmeP04},
finds two copies of each of the contributing Feynman diagrams calculated by one of the authors (MLW) in conjunction with Cho \cite{Cme04} and also Pak \cite{CmeP04},
the other two being quadratically divergent and therefore not contributing after renormalisation.

We can find the same result using $\zeta$-function renomalisation with appropriate considerations for causality.

We find therefore that the analysis in these two papers holds in this effective theory yielding a stable vacuum condensate at one loop.
}

\section{Application of Clairaut formalism to the Rotation-Boost decomposition of the gravitational connection} \label{sec:Clairaut}
\subsection{A review of the Hamiltonian-Clairaut formalism}
Here we review the main ideas and formulae of the Clairaut-type formalism for
singular theories \cite{D10,D11,dup2018e}. Let us consider a singular Lagrangian
$L\left(  q^{A},v^{A}\right)  =L^{\mathrm{deg}}\left(  q^{A},v^{A}\right)  $,
$A=1,\ldots n$, which is a function of $2n$ variables ($n$ generalized
coordinates $q^{A}$ and $n$ velocities $v^{A}=\dot{q}^{A}=dq^{A}/dt$) on the
configuration space $\mathsf{T}M$, where $M$ is a smooth manifold, for which
the Hessian's determinant is zero. Therefore, the rank of the Hessian matrix $W_{AB}%
=\tfrac{\partial^{2}L\left(  q^{A},v^{A}\right)  }{\partial v^{B}\partial
v^{C}}$ is $r<n$, and we suppose that $r$ is constant. We can rearrange the
indices of $W_{AB}$ in such a way that a nonsingular minor of rank $r$ appears
in the upper left corner. Then, we represent the index $A$ as follows: if
$A=1,\ldots,r$, we replace $A$ with $i$ (the \textquotedblleft
regular\textquotedblright\ or \textquotedblleft canonical\textquotedblright~index), and, if $A=r+1,\ldots,n$ we replace $A$
with $\alpha$ (the \textquotedblleft degenerate\textquotedblright\ or \textquotedblleft non-canonical\textquotedblright~index).
Obviously, $\det W_{ij}\neq0$, and $\operatorname{rank}W_{ij}=r$. Thus any set
of variables labelled by a single index splits as a disjoint union of two
subsets. We call those subsets regular (having Latin indices) and degenerate
(having Greek indices). Canonical DOFs are obviously described by the former of these subsets while other DOFs can be placed in the second if their contribution to the Wronskian vanishes. 
As was shown in \cite{D10,D11}, the \textquotedblleft
physical\textquotedblright\ Hamiltonian can be presented in the form%
\begin{equation} 
H_{phys}\left(  q^{A},p_{i}\right)  =\sum_{i=1}^{r}p_{i}V^{i}\left(
q^{A},p_{i},v^{\alpha}\right)  +\sum_{\alpha=r+1}^{n}B_{\alpha}\left(
q^{A},p_{i}\right)  v^{\alpha}-L\left(  q^{A},V^{i}\left(  q^{A}%
,p_{i},v^{\alpha}\right)  ,v^{\alpha}\right)  , \label{hph1}%
\end{equation}
where the functions%
\begin{equation}
B_{\alpha}\left(  q^{A},p_{i}\right)  \overset{def}{=}\left.  \dfrac{\partial
L\left(  q^{A},v^{A}\right)  }{\partial v^{\alpha}}\right\vert _{v^{i}%
=V^{i}\left(  q^{A},p_{i},v^{\alpha}\right)  } \label{h}%
\end{equation}
are independent of the unresolved velocities $v^{\alpha}$ since
$\operatorname{rank}W_{AB}=r$. Also, the r.h.s. of (\ref{hph1}) does
not depend on the degenerate velocities $v^{\alpha}$%
\begin{equation}
\dfrac{\partial H_{phys}}{\partial v^{\alpha}}=0, \label{hpv}%
\end{equation}
which justifies the term \textquotedblleft physical\textquotedblright. The
Hamilton-Clairaut system which describes any singular Lagrangian classical
system (satisfying the second order Lagrange equations) has the form%
\begin{align}
  \dfrac{dq^{i}}{dt}=&\left\{  q^{i},H_{phys}\right\}  _{phys}-\sum
_{\beta=r+1}^{n}\left\{  q^{i},B_{\beta}\right\}  _{phys}\dfrac{dq^{\beta}%
}{dt},\ \ i=1,\ldots r\label{q1}\\
  \dfrac{dp_{i}}{dt}=&\left\{  p_{i},H_{phys}\right\}  _{phys}-\sum
_{\beta=r+1}^{n}\left\{  p_{i},B_{\beta}\right\}  _{phys}\dfrac{dq^{\beta}%
}{dt},\ \ i=1,\ldots r\label{q2}\\
\sum_{\beta=r+1}^{n}&\left[  \dfrac{\partial B_{\beta}}{\partial q^{\alpha}%
}-\dfrac{\partial B_{\alpha}}{\partial q^{\beta}}+\left\{  B_{\alpha}%
,B_{\beta}\right\}  _{phys}\right]  \dfrac{dq^{\beta}}{dt}\nonumber\\
&  =\dfrac{\partial H_{phys}}{\partial q^{\alpha}}+\left\{  B_{\alpha
},H_{phys}\right\}  _{phys},\ \ \ \ \ \ \ \ \ \ \ \ \alpha=r+1,\ldots,n
\label{q3}%
\end{align}
where the \textquotedblleft physical\textquotedblright\ Poisson bracket (in
regular variables $q^{i}$, $p_{i}$) is%
\begin{equation}
\left\{  X,Y\right\}  _{phys}=\sum_{i=1}^{n-r}\left(  \frac{\partial
X}{\partial q^{i}}\frac{\partial Y}{\partial p_{i}}-\frac{\partial Y}{\partial
q^{i}}\frac{\partial X}{\partial p_{i}}\right)  . \label{xyp}%
\end{equation}

Whether the variables $B_{\alpha}\left(  q^{A},p_{i}\right)  $ have a
nontrivial effect on the time evolution and commutation relations is
equivalent to whether or not the so-called \textquotedblleft$q^{\alpha}$-field
strength\textquotedblright%
\begin{equation}
\mathcal{F}_{\alpha\beta}=\dfrac{\partial B_{\beta}}{\partial q^{\alpha}%
}-\dfrac{\partial B_{\alpha}}{\partial q^{\beta}}+\left\{  B_{\alpha}%
,B_{\beta}\right\}  _{phys} \label{f}%
\end{equation}
is non-zero. The reader is referred to references \cite{D10,D14,D11} for more details.

\subsection{The contribution of the Clairaut formalism}
\subsubsection{$q^\alpha$ curvature} \label{subsec:A2curvature}
Substituting in this notation, %of Section \ref{sec:A2}, 
the angles $\phi, \theta$ are seen, in parallel with our previously published analysis \cite{meD15}, to be degenerate DOFs with unresolved velocities.
Indeed, their contribution to both Lagrangian and Hamiltonian vanishes when their derivatives vanish.

We use the CDG decomposition in which the embedding of a dominant direction $U(1)$ is denoted by $\hat{n}$
which is defined by,
\begin{equation} \label{eq:npolar}
\hat{n}_\chi(x)\equiv\cos\theta(x) \sin\phi(x) \,_\chi\hat{e}_{1} +\sin\theta(x) \sin
\phi(x)\,_\chi\hat{e}_{2} +\cos\phi(x)\,_\chi\hat{e}_{3}.
\end{equation}
We note that the angles are $\phi,\theta$ are independent of $\chi$ for the reasons discussed after eqn (\ref{eq:generators}) and need not be labelled.
The following will prove useful:
\begin{align}	\label{eq:nphipolar} 
\sin\phi(x) \,_\chi\hat{n}_{\theta}(x)  &  \equiv\int dy^{4} \frac{d\hat{n}%
(x)}{d\theta(y)} =\sin\phi(x)\,(-\sin\theta(x)\,_\chi\hat{e}_{1}+\cos
\theta(x)\,_\chi\hat{e}_{2}), \nonumber\\
_\chi\hat{n}_{\phi}(x)  &  \equiv\int dy^{4} \frac{d\hat{n}_\chi(x)}{d\phi(y)}
=\cos\theta(x)\cos\phi(x)\,_\chi\hat{e}_{1} \nonumber \\
&\hspace{20mm}+\sin\theta(x)\cos\phi(x)\,_\chi\hat{e}%
_{2}-\sin\phi(x)\,_\chi\hat{e}_{3}.
\end{align}
For later convenience we note that
\begin{align} \label{eq:nphiphi}
_\chi\hat{n}_{\phi\phi}(x) = -_\chi\hat{n}(x),\; &
_\chi\hat{n}_{\theta\theta}(x) = -\sin\phi \;_\chi\hat{n}(x) - \cos\phi(x) \;_\chi\hat{n}_\phi(x)\;, \nonumber \\
&_\chi\hat{n}_{\theta\phi}(x) = 0,\;\;_\chi\hat{n}_{\phi\theta}(x) = \cos\phi(x) \;_\chi\hat{n}_\theta(x),
\end{align}
and that the vectors $_\chi\hat{n}=\,_\chi\hat{n}_{\phi}\times \,_\chi\hat{n}_{\theta}$
form an orthonormal basis of the internal space. 
Substituting the above into the Cho connection in eqn~(\ref{eq:vecV}) gives
\begin{align}
g \,_\chi\vec{C}_{\mu}(x)  &  =(\cos\theta(x) \cos\phi(x) \sin\phi(x)\partial_{\mu
}\theta(x) +\sin\theta(x)\partial\phi(x))\,_\chi\hat{e}_{1}\nonumber\\
& +(\sin\theta(x) \cos\phi(x) \sin\phi(x) \partial_{\mu}\theta(x) -\cos
\theta(x) \partial\phi(x))\,_\chi\hat{e}_{2} -\sin^{2}\phi(x)\partial_{\mu}%
\theta(x)\,_\chi\hat{e}_{3}\nonumber\\
&  =\sin\phi(x)\,\partial_{\mu}\theta(x)\,_\chi\hat{n}_{\phi}(x)-\partial_{\mu
}\phi(x)\,_\chi\hat{n}_{\theta}(x)
\end{align}
from which it follows that
\begin{equation}
g^{2} \,_\chi\vec{C}_{\mu}(x)\times \,_\chi\vec{C}_{\nu}(x) =\sin\phi(x)(\partial_{\mu}\phi(x)
\partial_{\nu}\theta(x) -\partial_{\nu}\phi(x)\partial_{\mu}\theta(x))\hat
{n}_\chi(x),\label{eq:CxC}%
\end{equation}
where we again see that higher order time derivatives are thwarted.

Since their Lagrangian terms do not fit the form of a canonical DOFs we consider them instead to be degenerate, having no canonical DOFs of their own but manifesting through their alteration 
of the EOMs of the dynamic variables.
Finding these alterations first requires the Clairaut-related quantities 
\begin{align}
B_{\phi}(x) & = \int dy^{3} \frac{\delta\mathcal{L}}%
{_{x}\partial_{0} \phi(x)}\nonumber\\
& =\sum_{\chi = R,L} \int dy^{3} \int dy_0 \, \delta(x_{0} - y_{0}) \Big( \sin\phi(y)
_{y}\partial_{\mu}\theta(y)\hat{n}_\chi(y) \nonumber \\
&\hspace{25mm}+\,_\chi\hat{n}_{\theta}(y)\times {}_\chi\vec{X}_{\mu
}(y)\Big)\cdot{}_\chi\vec{R}_{0\mu}(y)\, \delta^{3}(\vec{x}-\vec{y})\nonumber\\
& = \sum_{\chi = R,L}\Big(\sin\phi(x) \,\partial_{\mu}\theta(x)\hat{n}_\chi(x) +\,_\chi\hat{n}_{\theta
}(x)\times{}_\chi\vec{X}_{\mu}(x)\Big)\cdot{}_\chi\vec{R}_{0\mu}(x),\label{eq:gphi}\\
B_{\theta}(x)&= \int dy^{3} \frac{\delta\mathcal{L}%
}{_{x}\partial_{0} \theta(x)}\nonumber\\
&  =-\sum_{\chi = R,L} \int dy^{3} \int dy_{0} \delta(x_{0}-y_{0})\sin\phi(y) \Big( _{y}%
\partial_{\mu}\phi(y)\,\hat{n}_\chi(y) \nonumber\\
&\hspace{25mm}+\sin\phi(y)\,_\chi\hat{n}_{\phi}(y)\times{}_\chi\vec{X}_{\mu}(y)\Big)\cdot{}_\chi\vec{R}_{0\mu}(y) 
\,\delta^{3}(\vec{x}-\vec{y})\nonumber\\
& =-\sum_{\chi = R,L} \sin\phi(x) \Big(\partial_{\mu}\phi(x)\,\hat{n}_\chi(x)+ 
\,_\chi\hat{n}_{\phi}(x)\times{}_\chi\vec{X}_{\mu}(x)\Big)\cdot{}_\chi\vec{R}_{0\mu}(x). 
\label{eq:gtheta}%
\end{align}

\begin{align}
\frac{\delta B_{\phi}(x)}{\delta\theta(y)}=  
&  \sum_{\chi = R,L} \Big( \sin\phi(x)\,_\chi\hat{n}_{\theta\theta}(x) \times {}_\chi\vec{X}_\mu \cdot {}_\chi\vec{R}_{0\mu}(x) - \,_\chi T_\phi (x) \Big)
\delta^{4}(x-y), \\
\frac{\delta B_{\theta}(x)}{\delta\phi(y)}=  
&  -\sum_{\chi = R,L}\Big( \cos \phi(x)\Big(\partial_{\mu}\phi(x)\,\hat{n}_\chi(x)+\,_\chi\hat{n}_{\phi}(x)
\times{}_\chi\vec{X}_{\mu}(x)\Big)\cdot \Big( {}_\chi\vec{R}_{0\mu}(x) + \,_\chi\vec{H}_{0\mu}(x) \Big) \nonumber \\
& \hspace{90mm} + \,_\chi T_\theta (x) \Big)\delta^{4}(x-y), 
\end{align}
where
\begin{align}
_\chi T_\phi (x) = & \,\partial_k \Big[ \sin \phi(x) \, \hat{n}_\chi \cdot {}_\chi\vec{R}_{0k} (x)
- \Big( \sin \phi(x) \partial_k \theta (x) 
+ \,_\chi\hat{n}_\theta (x) \times {}_\chi\vec{X}_k \cdot \hat{n}_\chi \Big) \partial_0 \phi(x) \Big],  \\
_\chi T_\theta (x) = & - \partial_k \Big[ \sin \phi(x) \Big( \hat{n}_\chi \cdot {}_\chi\vec{R}_{0k} (x) 
+ \Big( \partial_k \phi (x) + \,_\chi\hat{n}_\phi (x) \times {}_\chi\vec{X}_k \cdot \hat{n}_\chi \Big) \partial_0 \theta(x)
\Big)\Big],
\end{align}
are the surface terms arising from derivatives 
$\frac{\delta (\partial \theta)}{\delta \theta},\,\frac{\delta (\partial \phi)}{\delta \phi}$ and the latin index $k$
is used to indicate that only spacial indices are summed over.

This yields the $q^{\alpha}$-curvature
\begin{align}
\label{eq:curvature}\mathcal{F}_{\theta\phi}(x) =&\int dy^{4} \Big( \frac
{\delta B_{\theta}(x)}{\delta\phi(y)}-\frac{\delta B_{\phi}(x)}{\delta
\theta(y)}\Big)\delta^{4}(x-y) +\{B_{\phi}(x),B_{\theta}(x)\}_{phys}%
\nonumber\\
=&-\sum_{\chi = R,L} \cos\phi(x)\Big(\partial_{\mu}\phi(x)\,\hat{n}_\chi(x)+ \,_\chi\hat{n}_{\phi}(x) \times
{}_\chi\vec{X}_{\mu}(x) \Big) \cdot \Big({}_\chi\vec{R}_{0\mu}(x) + \,_\chi\vec{H}_{0\mu}(x) \Big) \nonumber \\
&-\sum_{\chi = R,L} \sin\phi(x)\,_\chi\hat{n}_{\theta\theta}(x) \times {}_\chi\vec{X}_\mu(x) 
\cdot{}_\chi\vec{R}_{\mu 0}(x)  %\nonumber \\
  + \sum_{\chi = R,L}\Big( \,_\chi T_\phi (x) - \,_\chi T_\theta (x)\Big).
\end{align}
where we have used that the bracket $\{B_{\phi},B_{\theta}\}_{phys} $
vanishes because $B_{\phi}$ and $B_{\theta}$ share the
same dependence on the dynamic DOFs and their derivatives.

In earlier work on the Clairaut formalism \cite{D11,D10} this was called the
$q^{\alpha}$-field strength, but we call it $q^{\alpha}$-curvature in quantum
field theory applications to avoid confusion.

This non-zero $\mathcal{F}^{\theta\phi}$ is necessary, and usually
sufficient, to indicate a non-dynamic contribution to the conventional
Euler-Lagrange EOMs. More
significant is a corresponding alteration of the quantum commutators, with
repurcussions for canonical quantisation and the particle number.

%We shall neglect to prefix the symmetric field combinations with the suffix '$_+$' for the rest of this chapter for the sake of visual clarity, trusting that it shall 
%be understood in context since the antisymmetric combinations do not couple to the monopole background.

\subsubsection{Corrections to the equations of motion}
Generalizing eqs.~(7.1,7.3,7.5) in \cite{D10} (see also the discussion around eqn~(\ref{hph1})
%in appendix \ref{app:Clairaut}
\begin{equation}
\label{eq:alterEOM}\partial_{0}q(x)=\{q(x),H_{phys}\}_{new}= \frac{\delta
H_{phys}}{\delta p(x)} -\int dy^{4} \sum_{\alpha=\phi,\theta}
\frac{\delta B_{\alpha}(y)}{\delta p(x)} \partial_{0}\alpha(y),
\end{equation}
the derivative of the Abelian component, complete with corrections from the
monopole background is 
\begin{equation} \label{eq:altEOMc}
\partial_{0} \,_\chi c_{\sigma}(x) =\frac{\delta H_{phys}}{\delta\,_\chi\Pi^{\sigma}(x)}
-\int dy^{4} \sum_{\alpha=\phi,\theta} 
\frac{\delta B_{\alpha}(y)}{\delta\,_\chi\Pi^{\sigma}(x)}\partial_{0}\alpha(y).
\end{equation}

The effect of the second term is to remove the monopole contribution to  
$\frac{\delta H_{phys}}{\delta\,_\chi\Pi^{\sigma}}$. To see this, consider that, by construction,
the monopole contribution to the Lagrangian and Hamiltonian is dependent on the 
time derivatives of $\theta,\phi$, so the monopole component of 
$\frac{\delta H_{phys}}{\delta\,_\chi\Pi^{\sigma}}$ is
\begin{align}
\frac{\delta}{\delta\,_\chi\Pi_{\sigma}(x)} H_{phys} |_{\dot{\theta}\dot{\phi}} 
=& \frac{\delta}{\delta\,_\chi\Pi_{\sigma}(x)} \Big(
\frac{\delta H_{phys}}{\delta \partial_0 \theta(x)} \partial_0 \theta(x)
+ \frac{\delta H_{phys}}{\delta \partial_0 \phi(x)} \partial_0 \phi(x)
\Big) \nonumber \\
=& \frac{\delta}{\delta\,_\chi\Pi_{\sigma}(x)} \Big(
\frac{\delta L_{phys}}{\delta \partial_0 \theta(x)} \partial_0 \theta(x)
+ \frac{\delta L_{phys}}{\delta \partial_0 \phi(x)} \partial_0 \phi(x)
\Big) \nonumber \\
=& \frac{\delta}{\delta\,_\chi\Pi_{\sigma}(x)} \Big(
B_\theta (x) \partial_0 \theta (x) + B_\phi (x) \partial_0 \phi(x) \Big),
\end{align}
which is a consistency condition for eq.~(\ref{eq:altEOMc}). This confirms the necessity of
treating the monopole as a non-dynamic field.

We now observe that
\begin{equation}
\frac{\delta B_{\theta}(x)}{\delta \,_\chi c_{\sigma}(y)} = \frac{\delta B_{\phi}%
(x)}{\delta \,_\chi c_{\sigma}(y)} = 0,
\end{equation}
from which it follows that the EOMs of
$_\chi c_{\sigma}$ receives no correction. However its $\{,\}_{phys}$ contribution,
corresponding to the terms in the conventional EOM for the Abelian component,
already contains a contribution from the monopole field strength.

Repeating the above steps for the valence gluons $_\chi\vec{X}_{\mu}$, assuming $\sigma \ne 0$
and combining
\begin{equation}
\hat{D}_{0}\,_\chi\vec{\Pi}_{\sigma}(x) =\frac{\delta H}{\delta\,_\chi\vec{X}_{\sigma}(x)}
-\int dy^{4} \sum_{\alpha=\phi,\theta} 
\frac{\delta B_{\alpha}(y)}{\delta\,_\chi\vec{X}_{\sigma}(x)}\partial_{0}\alpha(y).
\end{equation}
with
\begin{align}
\frac{\delta B_{\phi}(y)}{\delta\,_\chi\vec{X}_{\sigma}(x)} =&-\Big(\Big(
\sin\phi(y)_{y}\partial_{\sigma}\theta(y)\hat{n}_\chi(y)
+\,_\chi\hat{n}_{\theta}(y)\times\,_\chi\vec{X}_{\sigma}(y)\Big) 
\times \,_\chi\vec{X}_0(y) \nonumber \\
& \hspace{55mm}-\,_\chi\hat{n}_\phi(y) \hat{n}_\chi \cdot \,_\chi\vec{R}_{0\sigma}(y) \Big)\delta^{4}(x-y),
\end{align}%
\begin{align}
\frac{\delta B_{\theta}(y)}{\delta\,_\chi\vec{X}^{\sigma}(x)} =&\Big(\Big(
\partial_{\sigma}\phi(y)\hat{n}(y)+\sin\phi(y)\, 
\hat{n}_{\phi}(y)\times\,_\chi\vec{X}_{\sigma}(y)\Big) 
\times \,_\chi\vec{X}_0(y) \nonumber \\
&\hspace{45mm}-\sin\phi(y) \,_\chi\hat{n}_{\theta}(y) \hat{n}_\chi(y) \cdot \,_\chi\vec{R}_{0\sigma}(y)\Big) \delta^{4}(x-y),
\end{align}
gives
\begin{align}
\hat{D}_{0}\,_\chi\vec{\Pi}_{\sigma}(x) =  &  
\frac{\delta H}{\delta\,_\chi\vec{X}_{\sigma}(x)} 
-\frac{1}{2}\Big(\Big(\sin\phi(x)(\partial_{\sigma}\phi(x)\partial_{0}%
\theta(x) -\partial_{\sigma}\theta(x)\partial_{0}\phi(x)\Big)\hat{n}_\chi(x)\nonumber\\
& +\Big(\sin\phi(x)\,_\chi\hat{n}_{\phi}(x)\partial_{0}\theta(x) -\,_\chi\hat{n}_{\theta}%
(x)\partial_{0}\phi(x)\Big) \times\,_\chi\vec{X}_{\sigma}(x)\Big)\times \,_\chi\vec{X}_0(x) \nonumber\\
=  &  \frac{\delta H}{\delta\,_\chi\vec{X}_{\sigma}(x)} -\frac{1}{2}
g^{2}\Big(\,_\chi\vec{C}_{\sigma}(x)\times\,_\chi\vec{C}_{0}(x) +\,_\chi\vec{C}_{0}(x)\times
\,_\chi\vec{X}_{\sigma}(x) \Big)\times\,_\chi\vec{X}_{0}(x).\label{eq:XEOM}%
\end{align}
This is the converse situation of the Abelian gluon, where 
their derivatives $_\chi\vec{X}_{\sigma}$
is uncorrected while their EOM
%$\frac{d\vec{\Pi}_{\sigma}}{dt}$
%, where $\vec{\Pi}_{\sigma}$ is the conjugate momentum of $\vec{X}_{\sigma}$
receives a correction which cancels the monopole's electric contribution to
$\{\hat{D}_{0}\,_\chi\vec{X}_{\sigma},H_{phys}\}_{phys}$. This is required by
the conservation of topological current.

\subsubsection{Corrections to the commutation relations} \label{subsubsec:commutation}
Corrections to the classical Poisson bracket correspond to corrections to the
equal-time commutators in the quantum regime. We shall see corrections for commutators with fields of different $SU(2)_\chi$ representations even though there were no such crossover terms 
in the effective potential calculation. 

Denoting conventional commutators as $[,]_{phys}$ and the corrected ones as $[,]_{new}$, for
$\mu,\nu\ne0$ we have
\begin{align} \label{eq:ccnew} 
[_\chi c_{\mu}(x),\,_{\tilde{\chi}} c_{\nu}(z)]_{new} =  [\,_\chi c_{\mu}(x),\,_{\tilde{\chi}} c_{\nu}(z)]_{phys}\hspace{65mm}\nonumber\\
-\int dy^{4}\Big( \frac{\delta B_{\theta}(y)}{\delta\,_\chi\Pi_{\mu}(x)}
\mathcal{F}_{\theta\phi}^{-1}(z)
\frac{\delta B_{\phi}(y)}{\delta\,_\chi\Pi_{\nu}(z)} - \frac{\delta B_{\phi}%
(y)}{\delta\,_{\tilde{\chi}}\Pi_{\mu}(x)} \mathcal{F}_{\phi\theta}^{-1}(z)
\frac{\delta B_{\theta}(y)}{\delta\,_\chi\Pi_{\nu}(z)}
\Big)  \delta^{4}(x-z)\nonumber\\
=   [\,_\chi c_{\mu}(x),\,_{\tilde{\chi}} c_{\nu}(z)]_{phys} \hspace{95mm}\nonumber\\
- \sin \phi(x) \sin\phi(z)( \partial_{\mu}%
\phi(x)\partial_{\nu}\theta(z)- \partial_{\nu}\phi(z)\partial_{\mu}\theta(x)) 
\mathcal{F}_{\theta\phi}^{-1}(z) \delta^{4}(x-z).
\end{align}
The second term on the final line, after integration over $d^4 z$,
clearly becomes
\begin{align}
H_{\mu\nu} (x) \sin\phi(x) \mathcal{F}_{\theta\phi}^{-1}(x),
\end{align}
indicating the role of the monopole condensate in the correction.
By contrast, the commutation relations 
\begin{align}
[\,_\chi c_{\mu}(x),\,_{\tilde{\chi}}\Pi_{\nu}(z)]_{new} = &\,[\, _\chi c_{\mu}(x),\,_{\tilde{\chi}}\Pi_{\nu}(z)]_{phys}, \nonumber \\
[\,_\chi\Pi_{\mu}(x),\,_{\tilde{\chi}}\Pi_{\nu}(z)]_{new} = &\,[\,_\chi\Pi_{\mu}(x),\,_{\tilde{\chi}}\Pi_{\nu}(z)]_{phys} ,
\end{align}
are unchanged. Nonetheless, the deviation from the canonical commutation shown in eqn~(\ref{eq:ccnew}) is inconsistent with the
particle creation/annihilation operator formalism of conventional second quantization, so that particle number is no longer 
well-defined for the $_\chi c_\mu$ fields.

Repeating for the valence part,
\begin{align} \label{eq:PiXnew}
[\,_\chi\Pi^a_{\mu}(x),&\,_{\tilde{\chi}}\Pi^b_{\nu}(z)]_{new} \newline \\
=& [\,_\chi\Pi^a_{\mu}(x),\,_{\tilde{\chi}}\Pi^b_{\nu}(z)]_{phys} 
-\int dy^{4} \Big( \frac{\delta B_{\theta}%
(y)}{\delta \,_\chi X^a_{\mu}(x)} \frac{\delta B_{\phi}(y)}{\delta \,_{\tilde{\chi}} X^b_{\nu}(z)} 
- \frac{\delta B_{\phi}(y)}{\delta X^a_{\mu}(x)} \frac{\delta B_{\theta
}(y)}{\delta \,_{\tilde{\chi}} X^b_{\nu}(z)} \Big) \,\mathcal{F}_{\theta\phi}^{-1}(z)\nonumber \\
=  &  [\,_\chi\Pi^a_{\mu}(x),\,_{\tilde{\chi}} Pi^b_{\nu}(z)]_{phys} 
+ \Big(\sin\phi(z) n^a_\phi(x) n^b_\theta(z) \,_\chi\vec{R}_{0\mu}(x) \cdot \hat{n}_\chi(x)
\,_{\tilde{\chi}}\vec{R}_{0\nu}(z) \cdot \hat{n}_{\tilde{\chi}}(z) \nonumber \\
&- \sin\phi(x) n^a_\theta(x) n^b_\phi(z) \,_\chi\vec{R}_{0\mu}(z) \cdot \hat{n}_\chi(z)
\,_{\tilde{\chi}}\vec{R}_{0\nu}(x) \cdot \hat{n}_{\tilde{\chi}}(x) \Big) 
\times \mathcal{F}_{\theta\phi}^{-1}(z)\,
\delta^{4}(x-z),
\end{align}
%where now $a,b$ are spatial indices and $\mathbf{\Pi}^a_{\mu}(x)$ is the $a^{th}$ component of $\mathbf{\vec{\Pi}}_{\mu}(x)$. 
where the second term on the final line, integrates over $d^4 z$ to become
\begin{align}
(n^a_\phi(x) n^b_\theta(x) - n^a_\theta(x) n^b_\phi(x))\sin\phi(x)
\,_\chi\vec{R}^{0\mu}(x) \cdot \hat{n}_\chi(x) \,_{\tilde{\chi}}\vec{R}^{0\nu}(x) \cdot \hat{n}_{\tilde{\chi}}(x)
\mathcal{F}_{\theta\phi}^{-1}(x),
\end{align}
while other relevant commutators are unchanged
\begin{align}
[\,_\chi X^a_{\mu}(x),\,_{\tilde{\chi}}\Pi^b_{\nu}(z)]_{new} = [\,_\chi X^a_{\mu}(x),\,_{\tilde{\chi}} \Pi^b_{\nu}(z)]_{phys}, \nonumber \\
[\,_\chi X^a_{\mu}(x),\,_{\tilde{\chi}} X^b_{\nu}(z)]_{new} = [\,_\chi X^a_{\mu}(x),\,_{\tilde{\chi}} X^b_{\nu}(z)]_{phys} .
\end{align}

\ignore{
\begin{align}
\frac{\delta B_{\theta}(y)}{\delta\vec{X}^{\sigma}(x)} =&\,\Big(\epsilon^{abc}\Big(
 \,_y\partial_{\sigma}\phi(y)\hat{n}_b(y) +\epsilon_{bde}\sin\phi(y)\,
\hat{n}^d_{\phi}(y)\times\vec{X}^e_{\sigma}(y)\Big)
\times \vec{X}_{c\,0}(y) \nonumber \\
&-\sin\phi \; \hat{n}^a_{\theta}(y) \mathbf{\hat{n}}(y) \cdot \mathbf{R}_{0\sigma}(y)\Big) \delta^{4}(x-y),
%(F_{0\sigma}(y) + \hatn(y) \cdot (\vec{X}_\mu(y) \times \vec{X}_0(y) - \vec{X}_\mu(y) \times \vec{X}_0(y) ) \delta^{4}(x-y).\nonumber\\ 
\end{align}
}

\ignore{
For the sake of completeness we expand the final term in these two equations, finding
\begin{align}
\mathbf{\hat{l}}(y) \cdot \mathbf{R}_{0\nu}(y) = &\,F_{0\sigma}(y) + \hatn(y) \cdot \Big(\vec{X}_\mu(y) \times \vec{X}_0(y) - \vec{X}_\mu(y) \times \vec{X}_0(y) \Big)
\end{align}
}

\section{Effective action} \label{sec:effective}
\subsection{Particle number and the monopole background} \label{subsec:operator}
It is textbook knowledge that gravitational curvature spoils canonical quantisation, but our approach gives a detailed
mechanism. It also provides some narrowly defined circumstances under which it may be salvaged.
%We now consider the specific circumstances under which these particle-spoiling corrections might vanish.
For monopole background $_\chi\vec{H}_{\mu\nu}$
the form of eqn (\ref{eq:ccnew}) indicates that they would arise for ${}_\chi c_\sigma$ polarised along either of the
$\mu,\nu$ directions.
% would lead to a correction which violates standard second quantisation and leaves particle number ill-defined.
The only way to avoid this is if ${}_\chi c_\sigma$ is polarised in
the direction of the monopole field strength, requiring that the Abelian component of the connection propagate at a right-angle to the monopole field-strength.
However, the form of the monopole field strength
%, given in eq~\ref{eq:}  are polarised
requires that a non-vanishing field must have a varying orientation in space, since it is proportional to the derivatives of the angles $\phi,\theta$. So even if the Abelian gauge component is
propagating at a right-angle to the monopole field-strength with its polarisation in the direction of the field strength, in general this could not be assumed to continue as the orientation of the
monopole field strength varied. However,
if the variation were gradual over space in comparison to the wavelength of ${}_\chi c_\mu$ then it might continue to propagate while adjusting to the required orientations in a manner analogous to
photon polarisation being rotated by successive, closely oriented, polarising filters.
On the other hand, if the wavelength of ${}_\chi c_\mu$ is significant compared to the length scale of the field variation then such a mechanism could not act and the particle's
energy would be either absorbed or deflected by the condensate, effectively suppressing the longer wavelengths and providing a measure of the background curvature.

One important observation is that the background field is (Lorentz) magnetic, so that at any point in spacetime a reference frame exists where the monopole field and its associated potential
lie entirely along the spatial directions.

The particle inconsistent contribution from eqn~(\ref{eq:PiXnew}) only occurs in the presence of a background electric component of the monopole field strength, vanishing when the
polarisation of ${}_\chi \vec{X}_\mu$ is orthogonal to the electric component of the background field. This restricts the polarisation for a transversally polarised field whose direction of propagation is 
not in the direction of this electric component, but not otherwise. Of course, the electric component of the background
monopole field can always be removed by a suitable Lorentz transformation, but this still leaves the particle interpretation frame-dependent.

Some authors have argued that the valence gluons in two-colour QCD gain an effective mass term \cite{KMS05,K06} via their quartic interaction with the non-trivial monopole condensate. 
A similar mechanism could apply to the valence components of this theory. Consider the following quartic term from eqs~(\ref{eq:QCD},\ref{eq:fieldstrength}),
\begin{align}
\frac{g^2}{4}& (_\chi\vec{C}_\mu(x) \times _\chi\vec{X}_\nu(x)) \cdot (_\chi\vec{C}^\mu(x) \times \,_\chi\vec{X}^\nu(x)) \nonumber \\
= \frac{g^2}{4}& (_\chi\vec{C}_\mu(x) \cdot \,_\chi\vec{C}^\mu(x) \; _\chi\vec{X}_\nu(x) \cdot \,_\chi\vec{X}^\nu(x)  
- \,_\chi\vec{C}_\mu(x) \cdot \,_\chi\vec{X}^\mu(x) \; _\chi\vec{X}_\mu(x) \cdot \,_\chi\vec{C}^\mu(x) ).
\end{align}
Remembering that the Lorentz monopole fields $_\chi\vec{C}_\mu$ have non-zero condensates yields the terms
\begin{equation}
\frac{g^2}{4} \langle_\chi\vec{C}_\mu(x) \cdot \,_\chi\vec{C}^\mu(x) \rangle \; \,_\chi\vec{X}_\nu(x) \cdot \,_\chi\vec{X}^\nu(x) ,
\end{equation}
so that the monopole condensate is seen to generate a mass term for the valence component.
Such a mass term is covariant under the gauge transformation because, as shown in the discussion of eqs~(\ref{eq:transform}), the valence components transform as sources
although explicitly adding a mass term for these fields would spoil renormalisability.  
In this case the valence components could also be longitudinally polarised. With longitudinal polarisation the only restriction is that the direction of propagation 
be orthogonal to the background electric component of the monopole field strength.
The valence component might therefore enjoy a limited particle interpretation under a range of circumstances.

We observe that the two monopole field strengths $_R\vec{H}_{\mu\nu}, \,_L\vec{H}_{\mu\nu}$ sum to give a net field strength lying purely along the rotation directions in the internal space.
Exactly how this affects the observed dynamics of the theory, or even if it does, is unclear. We were unable to find a linear combination of the gauge fields to separate rotation and boost generators
which was equivalent to the original theory. If there is an effect then a reasonable scenario is that the coupling 
to linear momentum would dominate that to rotational momentum at large distances, as determined by the length scale of the condensate.

\subsection{Hilbert-Einstein term} \label{subsec:EH}
Kim and Pak \cite{KP08} considered the
effects of a torsion condensate. They found the resulting background field strength, if constant, spontaneously generated an EH term if the curvature tensor is expanded around it
(see the discussion of eqn (45) in their paper \cite{KP08}). 

Since our background is attributable to an Abelian background field %has been shown to lie along the maximal Abelian subgroup 
we expect the effective theory to have an Abelianised EH term, similar to that derived by Cho \etal{}~\cite{COK12,COP15} when applying the CDG decomposition to the Levi-Civita tensor.
Such details must await further work, but we are 
encouraged to believe that the theory may be Wick rotated back to Lorentz space for a positive semi-definite effective theory. Not only do all quantum fields have kinetic terms with the correct sign, but 
the Lagrangian's lowest order derivative terms come from an emergent term sometimes added to rectify the non-semi positive definiteness.

\section{Discussion} \label{sec:discussion}
We have applied the CDG decomposition to a Lorentz gauge theory and confirmed that it has a monopole condensate at one loop. Using the Clairaut formalism we have found how the monopole background 
modifies the canonical EOMs for the physical DOFs. Lorentz gauge theory has the problem of being non-positive semi-definite, which can be handled by adding a EH term. We did not add such a term
but instead postponed the problem by Wick rotating the theory into Euclidean space where the Lorentz gauge group becomes locally isomorphic to $SU(2)_R \times SU(2)_L$. 
We found the spontaneous generation of a vacuum condensate which others have argued \cite{KP08,PKT12} leads to an effective Hilbert-Einstein term.

The CDG decomposition introduces an internal unit vector to indicate the local internal direction of the Abelian subgroup of the gauged symmetry group.
However, the unit vector used to specify this subgroup does not form a canonical EOM and is degenerate. 
If we expand it in terms of its angular dependence, since its information content is purely directional, then those angles are 
also degenerate and we do not derive canonical EOMs for them. They do however add additional terms with important consequences for the theory's physics. 
They may not be ignored therefore, but require appropriate theoretical tools to analyse them. The authors addressed these issues in a previous analysis of QCD.
The purpose of this paper was to do so for a theory relevant to gravity. The main advantages of working in a gauged Lorentz theory for us is that the
gauge fields have quadratic kinetic terms well-suited to our Clairaut-based approach in addition to the opportunity to apply analyses and even results from
$SU(2)$ Yang-Mills theories.

We have not considered the effects of matter fields in the fundamental representation. We do note in passing that differences in this part of the spectrum must lead to variations in the magnitude
for the monopole condensate, so the differences in their matter spectra suggest that this theory has significantly different infrared behaviour from that of $SU(2)$ QCD. 
We also observe that the net monopole condensate lies in a direction of a rotation generator. We have not been able to derive corresponding canonical DOFs to reflect this so the physical 
significance of this observation, if any, remains obscure.

We have left the inclusion of translation symmetry to subsequent work. 
A full gravitational theory must of course include the full Poincar\'{e} symmetry group, but we submit 
that our Lorentz-only theory makes a sufficiently good approximation to indicate some relevant phenomenology.
%comprise the essential DOFs concerning gravity.
%Furthermore, scalars are not affected in the more limited Lorentz symmetry only theory.

\appendix

\setcounter{section}{0} \setcounter{equation}{0}

\renewcommand{\thesection}{\Alph{section}}

\end{document}